\documentclass[superscriptaddress,amsmath,amssymb,prl,twocolumn]{revtex4}

\usepackage{setspace}
\usepackage{graphicx}  
\usepackage{dcolumn}   
\usepackage{bm}        
\usepackage{rotating}
\usepackage{hyperref}
\usepackage{xcolor}
\usepackage{adjustbox}
\usepackage{soul}

\newcommand{\sectionname}[1]{ \noindent{\bfseries #1}.---}

\begin{document}

\begin{flushleft}
KCL-PH-TH-2020-72
\\
CERN-TH-2021-060
\end{flushleft}


\title{Implications for First-Order Cosmological Phase Transitions from the Third LIGO-Virgo Observing Run
}
\author{Alba Romero}
\affiliation{Institut de F\'\i sica  d'Altes Energies (IFAE), Barcelona Institute of Science and Technology, E-08193 Barcelona, Spain}
\author{Katarina Martinovic}
\affiliation{Theoretical Particle Physics and Cosmology Group, \, Physics \, Department, \\ King's College London, \, University \, of London, \, Strand, \, London \, WC2R \, 2LS, \, United Kingdom}
\author{Thomas A. Callister}
\affiliation{Center for Computational Astrophysics, Flatiron Institute, New York, New York 10010, USA}
\author{Huai-Ke Guo}
\affiliation{Department of Physics and Astronomy, University of Oklahoma, Norman, Oklahoma 73019, USA}
\author{Mario Mart\'\i nez}
\affiliation{Institut de F\'\i sica  d'Altes Energies (IFAE), Barcelona Institute of Science and Technology, E-08193 Barcelona, Spain}
\affiliation{Catalan Institution for Research and Advanced Studies (ICREA), E-08010 Barcelona, Spain}
\author{Mairi Sakellariadou}
\affiliation{Theoretical Particle Physics and Cosmology Group, \, Physics \, Department, \\ King's College London, \, University \, of London, \, Strand, \, London \, WC2R \, 2LS, \, United Kingdom}
\affiliation{Theoretical Physics Department, CERN, Geneva, Switzerland}
\author{Feng-Wei Yang}
\affiliation{Department of Physics and Astronomy, University of Utah, Salt Lake City, Utah 84112, USA}
\author{Yue Zhao}
\affiliation{Department of Physics and Astronomy, University of Utah, Salt Lake City, Utah 84112, USA} 


\begin{abstract}
We place
constraints on the normalized energy density in gravitational waves 
from first-order strong phase transitions 
using data from Advanced LIGO and Virgo's first, second and third observing runs.
First, adopting a broken power law model, 
we place $95 \%$ confidence level upper limits simultaneously on the gravitational-wave energy density at 25 Hz from unresolved compact binary mergers, 
 $\Omega_{\rm CBC} < 6.1 \times 10^{-9}$ , and  
strong first-order phase transitions, $\Omega_{\rm BPL} < 4.4 \times 10^{-9}$. The inclusion of the former is necessary since we expect this astrophysical signal to be the foreground of any detected spectrum.
We then consider two more complex phenomenological models, limiting at 25 Hz the gravitational-wave background due to bubble collisions to $\Omega_{\rm pt} < 5.0\times 10^{-9}$ and the background due to sound waves to $\Omega_{\rm pt} < 5.8\times10^{-9}$ at $95 \%$ confidence level for phase transitions occurring at temperatures above $10^8$~GeV.

\vspace{0.5cm}

\noindent DOI: \href{https://doi.org/10.1103/PhysRevLett.126.151301}{10.1103/PhysRevLett.126.151301}
\end{abstract}



{\maketitle}

\sectionname{Introduction} The Advanced LIGO~\cite{2015AdvancedLIGO} and Advanced Virgo~\cite{Acernese_2014} detection of gravitational waves (GWs) from compact binary coalescences (CBCs)~\cite{Abbott:2020niy} offers a novel and powerful tool in understanding our universe and its evolution. We have detected CBCs, and before the detectors reach their designed sensitivity we may detect a stochastic gravitational-wave background (SGWB) produced by many weak, independent and unresolved sources of cosmological or astrophysical origin~\cite{Allen:1996vm,Maggiore:2000gv,Caprini:2018mtu}. Among the former, phase transitions occurring in the early universe, is one of the plausible mechanisms leading to a SGWB. 

The universe might have undergone a series of phase transitions (see, e.g., Refs.~\cite{Mazumdar:2018dfl,Hindmarsh:2020hop}). In the case of a first-order phase transition (FOPT), once the temperature 
drops below a critical value, the universe transitions from a meta-stable phase to a stable one, through 
a sequence of bubble nucleation, growth, and merger. During this process, a SGWB is expected to be 
generated~\cite{Witten:1984rs,Hogan:1986qda}.

Many compelling extensions of the standard model predict strong FOPTs, e.g., grand unification models~\cite{Croon:2019kpe,Okada:2020vvb,Huang:2020bbe}, supersymmetric models~\cite{Huber:2015znp,Garcia-Pepin:2016hvs,Bian:2017wfv,Demidov:2017lzf,Haba:2019qol,Craig:2020jfv}, extra dimensions~\cite{Yu:2019jlb,Megias:2020vek}, composite Higgs models~\cite{Davoudiasl:2017zws,Bruggisser:2018mrt,Bian:2019kmg,DeCurtis:2019rxl,Xie:2020bkl,Agashe:2019lhy,Huang:2020mso} and models with an extended Higgs sector (see, e.g., Refs.~\cite{Huang:2016cjm,Ramsey-Musolf:2019lsf}).
Generally there might exist symmetries beyond the ones of the standard model, which are
 spontaneously broken through a FOPT; for example the Peccei-Quinn symmetry~\cite{Hebecker:2016vbl,Dev:2019njv,vonHarling:2019gme,DelleRose:2019pgi,Ghoshal:2020vud}, the $B-L$ symmetry~\cite{Jinno:2016knw,Hasegawa:2019amx,Bian:2019szo,Chao:2017ilw}, or the left-right symmetry~\cite{Brdar:2019fur}. 
 The nature of cosmological phase transitions depends strongly on the particle physics model at high energy scales.

The SGWB sourced by a FOPT spans a wide frequency range. The peak frequency is mainly determined by the temperature 
$T_{\rm pt}$ at which the FOPT occurs. Interestingly, if  $T_{\rm pt}\sim(10^7 - 10^{10})$~GeV -- an energy scale not accessible by any 
existing terrestrial accelerators -- the produced SGWB is within  the frequency range  of  Advanced  LIGO and 
Advanced Virgo~\cite{Lopez:2013mqa,Dev:2016feu}.  Such an energy scale is well-supported by either the Peccei-Quinn axion 
model~\cite{Peccei:1977ur}, which solves the strong CP problem and provides a dark matter candidate, or 
high-scale supersymmetry models~\cite{Wells:2003tf,Arvanitaki:2012ps,ArkaniHamed:2012gw}, among others. 
Especially, for axionlike particles, the upper end of the $T_{\rm pt}$ we probe is at the energy scale where astrophysical constraints, such as stellar cooling, lose their sensitivities \cite{Hewett:2012ns}.  In addition,  the lower end of the $T_{\rm pt}$ fits well in minisplit SUSY models where the Higgs mass is explained.

The well-motivated SGWB search is performed by cross-correlating strain data from different GW detectors \cite{Allen:1996vm, Romano:2016dpx}. 
No SGWB signal has been observed in the last three observation periods (O1-O3) of the LIGO/Virgo/KAGRA Collaboration (LVKC)~\cite{O3_isotropic}.
Nevertheless, one can use the data to constrain the energy density of gravitational waves, and consequently the underlying particle physics models. This is the aim of this Letter.


\sectionname{SGWB from phase transitions} In a FOPT, it is  well established that GW can be produced by mainly three sources: bubble collisions, sound waves, and magnetohydrodynamic turbulence (see, e.g., Refs.~\cite{Caprini:2015zlo,Cai:2017cbj,Weir:2017wfa,Hindmarsh:2020hop} for recent reviews). The GWs thus 
produced is a SGWB, described by the
energy density spectrum:
$\Omega_{\rm GW}(f) = {d\rho_{\rm GW}}/(\rho_{\rm c} {d \ln f})$ with $\rho_c$ the present critical energy density $\rho_{\rm c}=3c^2 H_0^2/(8\pi G)$. Each spectrum 
can be well approximated by a broken power law, with its peak frequency 
determined by the typical length scale at the transition, 
the mean bubble separation $R_{\text{pt}}$ which is related to the inverse time duration of the transition $\beta$, and also by the amount of redshifting determined by $T_{\text{pt}}$ and the cosmic history. The amplitude of each contribution is largely determined by the energy released normalized by the radiation energy density $\alpha$, its fraction
going into the corresponding source and the bubble wall velocity $v_{\rm w}$.
Here we do not consider the contribution from magnetohydrodynamic turbulence as it always
happens together with sound waves and is subdominant. In addition, we note that
its spectrum is the least
understood and might witness significant changes in the future~\cite{Caprini:2015zlo,Kahniashvili:2008pf,Kahniashvili:2008pe,Kahniashvili:2009mf,Caprini:2009yp,Kisslinger:2015hua,Pol:2019yex}.

{The dominant source for GW production in a thermal transition, as most commonly encountered in the early universe,
is the sound waves in the plasma
induced by the coupling between the scalar field and the
thermal bath~\cite{Hindmarsh:2015qta,Hindmarsh:2013xza,Hindmarsh:2017gnf}.}
A good analytical understanding of
this spectrum has been achieved through the sound shell model~\cite{Hindmarsh:2016lnk,Hindmarsh:2019phv,Guo:2020grp}, though it still does not capture all the 
physics~\cite{Hindmarsh:2017gnf,Cutting:2019zws,Hindmarsh:2020hop} to match perfectly the result from numerical simulations~\cite{Caprini:2015zlo,Hindmarsh:2015qta}. 
We use the spectrum from
numerical simulations:
{
\begin{eqnarray}
\Omega_{\textrm{sw}}(f)h^{2}=2.65\times10^{-6}\left( \frac{H_{\rm pt}}{\beta}\right) \left(\frac{\kappa_{\rm sw} \alpha}{1+\alpha} \right)^{2} 
\left( \frac{100}{g_{\ast}}\right)^{1/3} \nonumber \\
 \times v_{\rm w} \left(\frac{f}{f_{\text{sw}}} \right)^{3} \left( \frac{7}{4+3(f/f_{\textrm{sw}})^{2}} \right) ^{7/2}  \Upsilon(\tau_{\text{sw}})\ , \quad 
\label{eq:soundwaves}
\end{eqnarray}
}
where $\kappa_{\rm sw}$ is the fraction of vacuum energy converted into the kinetic energy of the bulk flow, a function of $v_{\rm w}$ and $\alpha$~\cite{Espinosa:2010hh,Giese:2020rtr};
$H_{\rm pt}$ is the Hubble parameter at $T_{\text{pt}}$; $g_{\ast}$ is the number of relativistic degrees of freedom, chosen to be 100 in our analysis; $h$ is the dimensionless Hubble parameter;
$f_{\textrm{sw}}$ is the present peak frequency,
{
 \begin{align}
f_{\textrm{sw}}=19\frac{1}{v_{\rm w}}\left(\frac{\beta}{H_{\rm pt}} \right) \left( \frac{T_{\rm pt}}{100\textrm{GeV}} \right) \left( \frac{g_{\ast}}{100}\right)^{\frac{1}{6}} \mu\textrm{Hz} ,
\end{align}
}
and $\Upsilon= 1 -(1 + 2 \tau_{\text{sw}} H_{\rm pt})^{-1/2}$~\cite{Guo:2020grp} which is a suppression factor due to the finite lifetime~\cite{Guo:2020grp,Ellis:2020awk}, $\tau_{\text{sw}}$, of sound waves. $\tau_{\text{sw}}$ is typically smaller
than a Hubble time unit~\cite{Ellis:2019oqb,Caprini:2019egz} and is usually chosen to be the timescale for the onset of turbulence~\cite{Weir:2017wfa}, $\tau_{\text{sw}} \approx {R_{\rm pt}}/{\bar{U}_f}$, with $R_{\rm pt} = (8\pi)^{1/3} v_{\rm w}/\beta$ for an exponential nucleation of bubbles~\cite{Hindmarsh:2019phv,Guo:2020grp}, and $\bar{U}_f^2 = 3 \kappa_{\rm sw} \alpha/[4(1+\alpha)]$~\cite{Weir:2017wfa}.

{
When sound waves, and thus also magnetohydrodynamic turbulence, are highly suppressed or absent,
bubble collisions can become dominant, e.g., for a FOPT in vacuum
of a dark sector which has no or very weak interactions with the standard plasma.
}
The resulting GW spectrum can be well modeled with the envelope approximation~\cite{Kosowsky:1992vn,Kosowsky:1992rz,Jinno:2016vai}, which assumes an infinitely thin bubble wall and neglects the contribution from overlapping bubble segments.
In the low-frequency regime, $\Omega_{\rm GW}\propto f^{3}$ 
from causality~\cite{Maggiore:2018sht}, and for high-frequencies $\Omega_{\rm GW}\propto f^{-1}$~\cite{Huber:2008hg} due to the dominant single bubble contribution as revealed by the analytical calculation~\cite{Jinno:2016vai}.
The spectrum is~\cite{Huber:2008hg,Jinno:2016vai,Weir:2017wfa} 
{
\begin{align}
\Omega_{\rm coll}(f) h^2 = 1.67\times 10^{-5} \Delta\left({H_{\rm pt} \over \beta } \right)^2 
\left( {   \kappa_{\phi} \alpha \over 1+ \alpha }\right)^2
 \nonumber \\
 \times
\left( { 100 \over g_{\ast}} \right)^{1/3}
 S_{\text{env}}(f), 
 \label{eq:bubbles}
\end{align}
}
where $\kappa_\phi=\rho_\phi/\rho_{\rm vac}$ denotes the fraction of vacuum energy converted into gradient energy of the scalar field. 
The amplitude $\Delta$ is $\Delta (v_{\rm w}) = 0.48 v_{\rm w}^3 /(1 + 5.3 v_{\rm w}^2 + 5 v_{\rm w}^4)$ and the spectral shape is 
$S_{\text{env}} = 1/(c_l \tilde{f}^{-3}  + (1-c_l-c_h) \tilde{f}^{-1}  + c_h \tilde{f})$
where $c_l = 0.064$, $c_h = 0.48$ and $\tilde{f}=f/f_{\text{env}}$ with $f_{\text{env}}$ the present peak frequency
\begin{eqnarray}
f_{\text{env}} =
16.5 
\left(\frac{f_{\rm bc}}{\beta}\right)
\left(\frac{\beta}{H_{\rm pt}}\right)
\left(\frac{T_{\rm pt}}{100\ \text{GeV}}\right)\left(\frac{g_{\ast}}{100}\right)^{\frac{1}{6}} \mu\text{Hz},
\end{eqnarray}
and
$f_{\rm bc}$ the peak frequency right after 
the transition ${f_{\rm bc}} = {0.35 \beta}/(1+0.069 v_{\rm w} + 0.69 v_{\rm w}^4)$. More recent simulations
going beyond the envelope approximation
show a steeper shape $f^{-1.5}$ for high frequencies~\cite{Cutting:2018tjt}, and it also varies from $f^{-1.4}$ to $f^{-2.3}$ as the wall thickness increases~\cite{Cutting:2020nla} (see also Refs.~\cite{Lewicki:2020jiv,Lewicki:2020azd,Di:2020nny}). 


\sectionname{Data Analysis} 
Here we take two analysis approaches.
First, we consider an approximated broken power law including main  
features of the shape and its peak. We then consider the phenomenological models Eqs. (\ref{eq:bubbles}) and (\ref{eq:soundwaves}), for contributions from bubble collisions and sound waves. 

{\sl I. Broken power law model}:
The spectrum can be approximated by
a broken power law (BPL) as
{
\begin{equation}
    \Omega_{\rm bpl}(f) =
              \Omega_*~\Big(\frac{f}{f_*}\Big)^{n_1}~ \Bigg[1+\Big(\frac{f}{f_*}\Big)^{\Delta}\Bigg]^{(n_2-n_1)/\Delta}.
\end{equation}
}
 Here $n_1=3$, from causality, and $n_2$ takes the values $-4$ and $-1$, for sound waves and bubble collisions, respectively. We fix 
the $n_1$ parameter in our search, but we let $n_2$ vary uniformly between -8 and 0, allowing for the values motivated by both contributions. The value for $\Delta$ is set to 2 for sound waves and 4 for approximating bubble collisions.  We run a Bayesian search for both values, but present results only for $\Delta=2$, since 
  it gives more conservative upper limits. 

 We follow Refs.~\cite{PhysRevLett.109.171102,PhysRevX.7.041058,PhysRevD.102.102005} to perform a Bayesian search and model selection.
 In addition to a search for the broken power law, we
undertake a study on simultaneous estimation of a CBC background and a broken power law background, because current estimates of the CBC background~\cite{Abbott:2017xzg,O3_isotropic} show it as a non-negligible component of any SGWB signal. The CBC background is very well approximated by an $f^{2/3}$ power law~\cite{PhysRevX.6.031018}. The challenge then is to search for a broken power law in the presence of a CBC background.
 
The log-likelihood for a single detector pair is Gaussian,
{
\begin{align}
    \log p(\hat C_{IJ}(f) | \boldsymbol{\theta}_{\rm gw}, \lambda) \propto -\frac{1}{2}\sum_{f}\frac{\left[\hat C_{IJ}(f) - \lambda \, \Omega_{\rm gw}(f, \boldsymbol{\theta}_{\rm gw}) \right]^2}{\sigma_{IJ}^2(f)} , 
    \label{eq:pe_ms:likelihood}
\end{align}
}
where $\hat C_{IJ}(f)$ and $\sigma_{IJ}(f)$ are data products of the analysis: $\hat{C}_{IJ}(f)$ is the cross-correlation estimator of the SGWB calculated using data from detectors $I$ and $J$, and $\sigma^2_{IJ}(f)$ is its variance~\cite{PhysRevD.59.102001}.
The search for an isotropic stochastic signal shows  no evidence of  correlated
magnetic noise, and a pure Gaussian noise model is still preferred by the data~\cite{O3_isotropic}. Therefore,  
here, a contribution from Schumann resonances~\cite{Thrane:2013npa, Cirone:2018guh, PhysRevD.102.102005} is neglected.
The model we fit to the data is $\Omega_{\rm GW}(f, \bm \theta_{\rm GW})$, with parameters $\boldsymbol\theta_{\rm GW}$. The parameter $\lambda$ captures calibration uncertainties of the detectors \cite{Sun:2020wke} and is marginalized over \cite{Whelan:2012ur}. For a multibaseline study, we add all log-likelihoods of individual baselines.
The set of GW parameters depends on the type of search we perform.  

The CBC spectrum is modeled as   
{
\begin{align}
    \Omega_{\rm cbc} = \Omega_{\rm ref} (f/f_{\rm ref})^{2/3},
\end{align}
}

\noindent
with $f_{\rm ref}=25~\rm{Hz}$. We consider three separate scenarios: contributions from unresolved CBC sources, with $\boldsymbol \theta_{\rm GW} = (\Omega_{\rm ref})$; broken power law contributions, with $\boldsymbol \theta_{\rm GW} = ({\Omega_*,f_*,n_2})$; and the combination of CBC and broken power law contributions, for which $\boldsymbol \theta_{\rm GW} = ({\Omega_{\rm ref}, \Omega_*, f_*, n_2})$. The priors used are summarized in Table \ref{tab:priors}. 
To compare GW models and assess which provides a better fit, we use ratios of evidences, otherwise known as Bayes factors. In particular, we consider  
$\log\cal{B}^{\rm CBC+BPL}_{\rm noise}$ and  $\log\cal{B}^{\rm CBC+BPL}_{\rm CBC}$ as indicative detection statistics.

{
\begin{table}[h]
\begin{center}
\begin{footnotesize}
\begin{tabular}{c| c} \hline
\multicolumn{2}{c}{\bf{Broken power law model}} \\ \hline
    {Parameter} & {Prior}\\
    \hline
    $\Omega_{\rm ref}$ & LogUniform($10^{-10}$, $10^{-7}$)\\
    $\Omega_*$  & LogUniform($10^{-9}$, $10^{-4}$)\\
    $f_*$ & Uniform(0, 256 Hz)\\
    $n_1$ & 3\\
    $n_2$ & Uniform(-8,0)\\
    $\Delta$ & 2\\
    \hline
    \hline
    \multicolumn{2}{c}{\bf{Phenomenological model}} \\ \hline
        {Parameter} & {Prior}\\
    \hline
        $\Omega_{\rm ref}$ & LogUniform($10^{-10}$, $10^{-7}$)\\
        $\alpha$ & LogUniform ($10^{-3}$, $10$)\\
        $\beta/H_{\rm pt}$ & LogUniform ($10^{-1}$, $10^{3}$)\\
        $T_{\rm pt}$ & LogUniform ($10^{5}$, $10^{10}$ GeV)\\
        $v_{\rm w}$ & 1 \\
        $\kappa_\phi$ & 1 \\
        $\kappa_{\rm sw}$ & $f(\alpha, v_{\rm w}) \in [0.1 - 0.9]$\\
        \hline
  \end{tabular}
  \end{footnotesize}
  \caption{List of prior distributions used for all parameters in the various searches. The narrow, informative prior on $\Omega_{\rm ref}$ stems from estimates of the CBC background \cite{Abbott:2017xzg}, and encompasses uncertainties on the mass and redshift distributions of CBCs~\cite{O3_isotropic,TheLIGOScientific:2016wyq}. The frequency prior is uniform across the frequency range considered since we have no further information about it.
}
  \label{tab:priors}
  \end{center}
\end{table}
}

{\sl II. Phenomenological model}:
Two scenarios are considered,  corresponding to dominant contributions from bubble collisions or sound waves, respectively, following 
an approach similar to Ref.~\cite{vonHarling:2019gme}. 
The analysis procedure follows closely that of the broken power law search, with $\boldsymbol \theta_{\rm GW} = ({\Omega_{\rm ref}, \alpha, \beta/H_{\rm pt}, T_{\rm pt} })$
including CBC background $\Omega_{\rm CBC}$,  and $\Omega_{\rm GW}$ from 
bubble collisions and sound waves described by Eqs.~(\ref{eq:bubbles}) and ~(\ref{eq:soundwaves}), respectively. 

For bubble collisions, $v_{\rm w}$ and $\kappa_\phi$ are set to unity.  The remaining model parameters are varied in the ranges in Table~\ref{tab:priors}.  We note that the GW spectra in Eqs.~(\ref{eq:bubbles}) and ~(\ref{eq:soundwaves}) may not be applicable when $\alpha\gtrsim 10$, and also a large $\alpha$ does not translate into a significant 
increase in the GW amplitude. Moreover, $\beta/H_{\text{pt}}$ is related to the mean bubble separation, up to an $O(1)$ coefficient, and one should be cautious when it is smaller than $1$~\cite{Ellis:2018mja,Ellis:2019oqb}. In this study, we conservatively choose $\beta/H_{\text{pt}}$ to be larger than 0.1.

For sound waves,  we initially set $v_{\rm w} = 1$, and then 
explore different values for  $v_{\rm w}$ in the range (0.7 - 1.0), corresponding to various detonation and hybrid modes of
fluid velocity profile~\cite{Cutting:2019zws,Espinosa:2010hh}.  
Here $\kappa_{\rm sw}$ is a function of $\alpha$ and $v_{\rm w}$, e.g., for  $v_{\rm w} = 1$, $\kappa_{\rm sw}$ increases from 0.1 to 0.9 as $\alpha$ increases from 0.1 to 10. 
The  rest of the parameters are varied as in the case of bubble collisions. 

{
\begin{figure}
    \centering
    \includegraphics[width=0.49\textwidth]{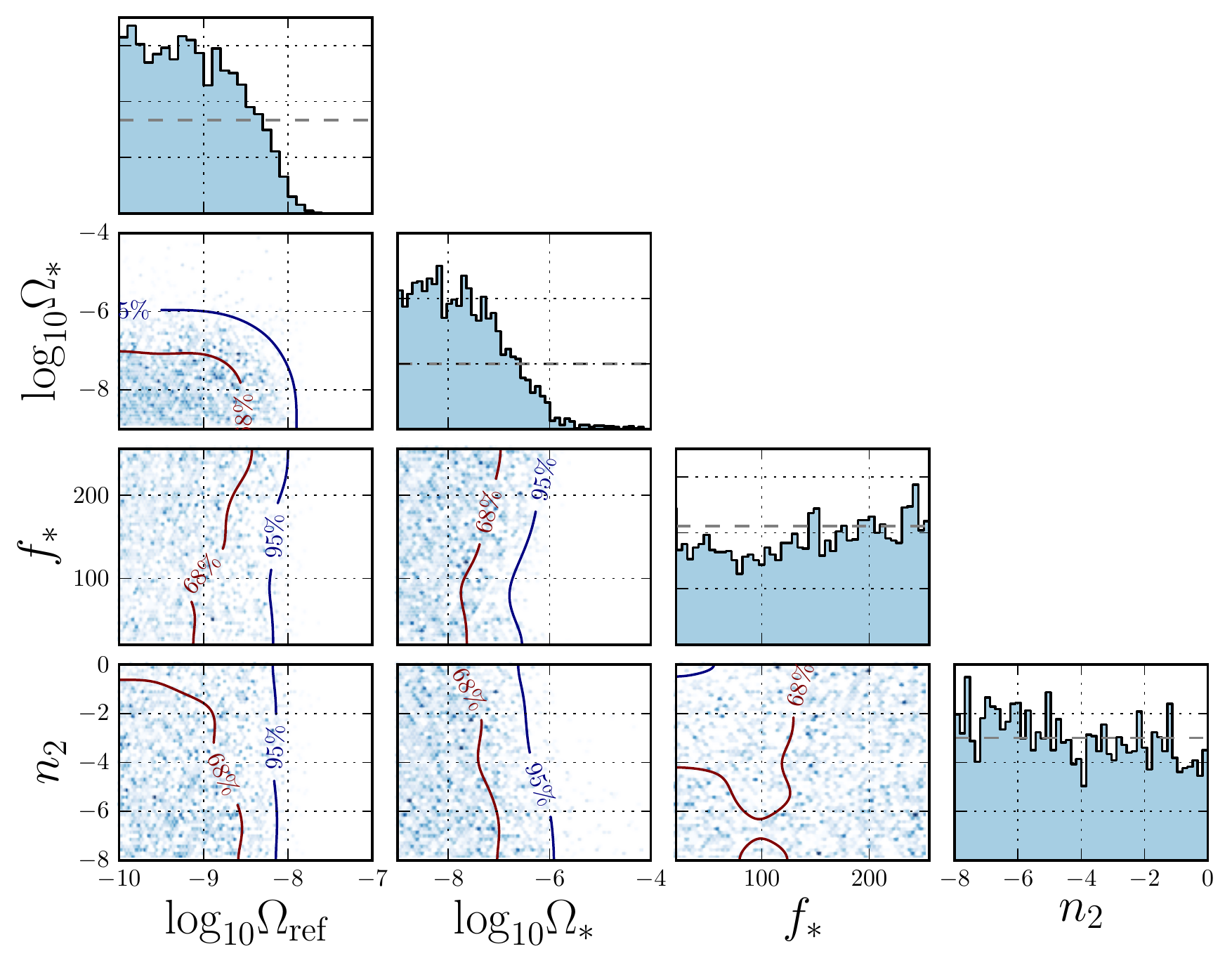}
    \caption{
Posterior distributions for the combined CBC and broken power law search as a function of 
$\log \Omega_{\rm ref}$ and the different parameters of the model.  The 68$\%$ and $95\%$ CL exclusion contours are shown. The horizontal dashed line in the 
posteriors indicate the flat priors used in the analysis. 
}
    \label{fig:corner_plot}
\end{figure}
}
\sectionname{Results} {\sl I. Broken power law model}:
In Fig.~\ref{fig:corner_plot} we present posterior distributions of parameters in the combined CBC and BPL search. The Bayes factor is $\log{\cal{B}}^{\rm CBC+BPL}_{\rm noise} = -1.4$, demonstrating no evidence of such a signal in the data from the three observing runs. The 2-d posterior of $\Omega_{\rm ref}$ and $\Omega_*$ allows us to place simultaneous estimates on the amplitudes of the two spectra. The 95\% confidence level (CL) upper limits are $\Omega_{\rm ref}=6.1 \times 10^{-9}$ and $\Omega_*=5.6 \times 10^{-7}$, respectively. 
If we take individual posterior samples of $\Omega_*$, $f_*$ and $n_2$ from Fig.~\ref{fig:corner_plot}, and combine them to construct a posterior of $\Omega_{\rm BPL}$, we estimate at 95\% CL $\Omega_{\rm BPL}(25\,\rm Hz) = 4.4 \times 10^{-9}$.
The width of the $n_2$ posterior suggests no preference for a particular value by the data, and we are unable to rule out any part of the parameter space at this time. Other searches give Bayes factors $\log{\cal{B}}^{\rm BPL}_{\rm noise}= -0.78$ and $\log{\cal{B}}^{\rm CBC+BPL}_{\rm CBC}= -0.81$, once again giving no evidence for a BPL signal, with or without CBCs considered.

To demonstrate the dependence of GW amplitude constraints on other parameters, we present 95\% CL upper limits on $\Omega_*$ for a set of $n_2$ and $f_*$ in Table~\ref{tab:bplproj}. We choose 
representative values of $n_2$, for bubble collisions, $n_2=$ -1 and -2, and for sound waves, $n_2=$ -4. The $f_*$ values are chosen to represent broken power laws that peak before, at, and after the most sensitive part of the LIGO-Virgo band, $f_*=\,25 \, \rm{Hz}$. As expected, the most constraining upper limits are obtained for a signal that peaks at  25 Hz. For the signal in the first column that peaks at 1 Hz, the faster it decays, the weaker it is at 25 Hz. Therefore, the more negative $n_2$ values give less constraining upper limits on the amplitude.  Finally, the signal that peaks at 200 Hz gives similar $\Omega_*$ upper limits for all values of $n_2$ since it resembles a simple $n_1=3$ power law in the range with largest SNR.
Note the upper limits in Table~\ref{tab:bplproj} are fundamentally different from results in Fig.~\ref{fig:corner_plot}. In the former case we fix $f_*$ and $n_2$ and find $\Omega_*^{95\%}$, while in the latter we marginalize over all parameters to obtain $\Omega_*^{95\%}$.

{
\begin{table}[h]
\begin{center}
\begin{footnotesize}
\begin{tabular}{c|c|c|c} \hline
\multicolumn{4}{c}{\bf{Broken power law model}} \\ \hline
      & $f_*=1\,\rm{Hz}$ & $f_*=25\, \rm{Hz}$ & $f_*=200\, \rm{Hz}$ \\
    \hline
    $n_2=-1$  & $3.3 \times 10^{-7}$ & $3.5 \times 10^{-8}$ & $2.8 \times 10^{-7}$  \\
    $n_2=-2$  & $8.2 \times 10^{-6}$ & $6.0 \times 10^{-8}$ & $3.7 \times 10^{-7}$ \\
    $n_2=-4$ & $5.2 \times 10^{-5}$ & 1.8 $\times 10^{-7}$ & $3.7 \times 10^{-7}$ \\
    \hline
     \end{tabular}
  \end{footnotesize}
  \caption{Upper limits for the energy density amplitude, $\Omega_*^{95\%}$, in the broken power law model for fixed values of the peak frequency, $f_*$, and negative power law index, $n_2$.}
  \label{tab:bplproj}
  \end{center}
\end{table}
}


{\sl II. Phenomenological model}:
We now estimate 95$\%$ CL  upper limits on 
$\Omega_{\rm coll}$ and $\Omega_{\rm sw}$   
from   
bubble collisions and sound waves respectively.  
The Bayesian analysis is 
repeated separately for $\Omega_{\rm coll}$ and $\Omega_{\rm sw}$ contributions, with  
priors stated in Table~\ref{tab:priors}, leading to Bayes factors $\log\cal{B}^{\rm CBC+coll}_{\rm noise}$= -0.74 and 
$\log\cal{B}^{\rm CBC+sw}_{\rm noise}$= -0.66, respectively. 

%
%

In Fig.~\ref{fig:BC} we present exclusion regions as a function of the different parameters of the CBC+FOPT model, now under the assumption 
that contributions from bubble collisions dominate,  with $v_{\rm w} = 1$ and $\kappa_\phi = 1$. 
In general, with the chosen prior, the data can exclude part of the parameter space at 95$\%$ CL, especially when $T_{\rm pt} > 10^8$ GeV, $\alpha > 1$, or $\beta/H_{\rm pt} < 1$.

{
\begin{figure}[htb]
\begin{center}
\includegraphics[width=0.49\textwidth]{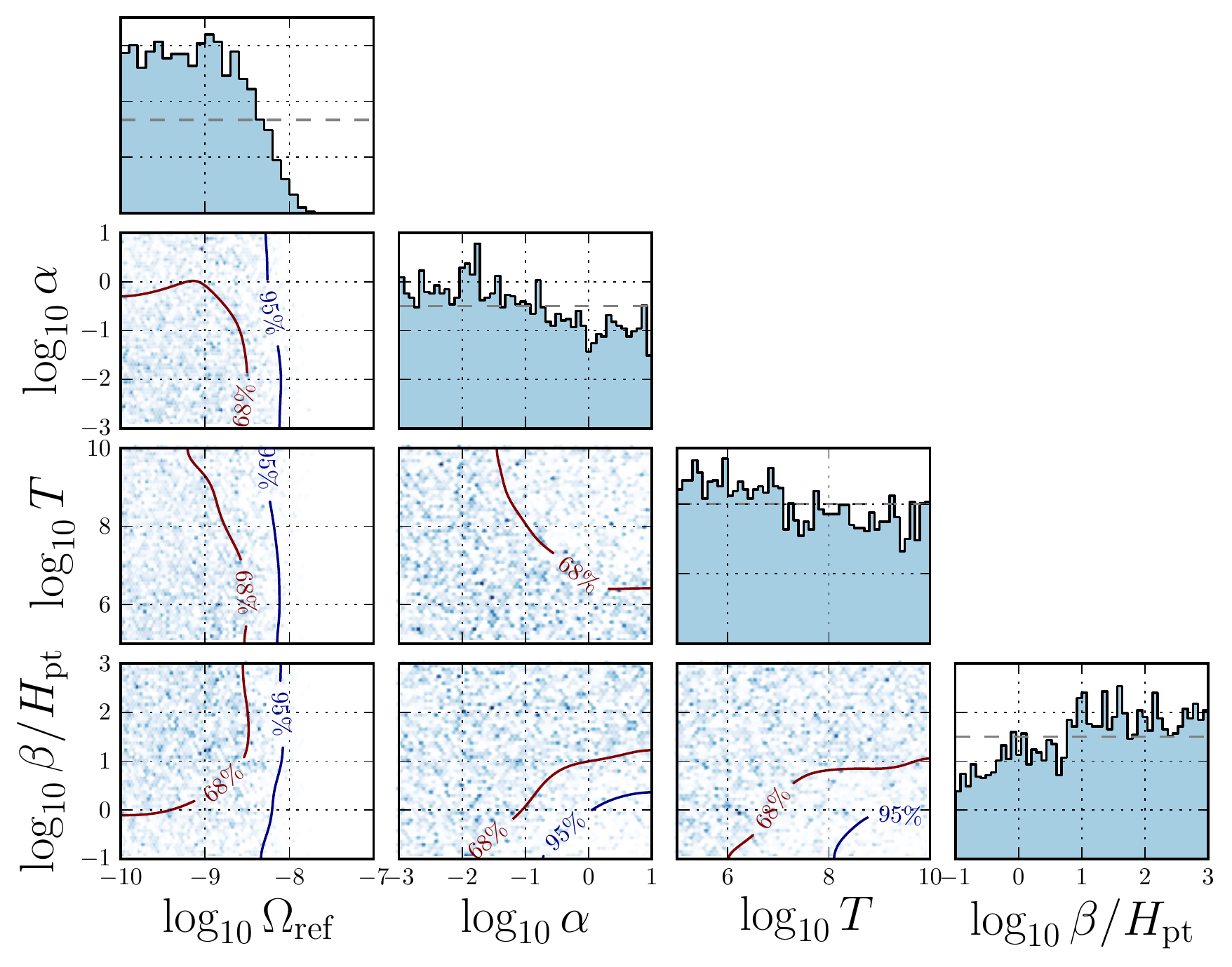}
\end{center}
\caption{\small
Posterior distributions for the CBC+FOPT search in the case of a phenomenological model with dominant bubble collision contributions as a function of 
$\log \Omega_{\rm ref}$ and the different parameters of the model.  The 68$\%$ and $95\%$ CL exclusion contours are shown. The horizontal dashed line in the 
posteriors indicate the flat priors used in the analysis. 
}
\label{fig:BC}
\end{figure}
}

Table~III presents 95$\%$ CL upper limits on $\Omega_{\rm coll}$(25 Hz) for several $\beta/H_{\rm pt}$ and $T_{\rm pt}$, where $\alpha$ is left as a 
free parameter to be inferred from the data. We consider three values for  $\beta/H_{\rm pt}$, namely 0.1, 1, and 10,  and four for $T_{\rm pt}$: $10^{7}$, $10^8$, $10^9$, and $10^{10}$~GeV. Our constraints on $\Omega_{\rm coll}$(25 Hz), as computed at the reference frequency of 25~Hz, vary in the range $4.0\times 10^{-9}$ to $ 1.0\times 10^{-8}$, with more stringent limits at large $\beta/H_{\rm pt}$ or large $T_{\rm pt}$. At the largest values of $\beta/H_{\rm pt}$ and $T_{\rm pt}$ there is not enough sensitivity to place constrains to the model.   
In all cases, the inferred upper limits on the CBC background range between $\Omega_{\rm ref}$ = $5.3\times 10^{-9}$ and $6.1\times 10^{-9}$.

{
\begin{table}[h]
       \begin{center}
               \begin{footnotesize}
                        \begin{tabular}{c|c|c|c|c} \hline
                                \multicolumn{5}{c}{\bf{Phenomenological model (bubble collisions)} }\\ \hline
                                \multicolumn{5}{c}{{$\Omega_{\rm coll}^{95\%}$ (25 Hz)}} \\ \hline 
                                 $\beta/H_{\rm pt}  \setminus  T_{\rm pt}$ & $10^7$ GeV &  $10^8$ GeV & $10^9$ GeV & $10^{10}$ GeV\\ \hline
                                 0.1  & $9.2\times10^{-9}$& $8.8\times10^{-9}$ &$1.0\times10^{-8}$ & $7.2\times10^{-9}$ \\
                                 1    & $1.0\times10^{-8}$& $8.4\times10^{-9}$ &$5.0\times10^{-9}$ & $-$\\ 
                                 10    & $4.0\times10^{-9}$& $6.3\times10^{-9}$ &$-$ &$-$ \\ \hline
                        \end{tabular}
                        \caption{
                        The 95$\%$ CL upper limits  on $\Omega_{\rm coll}^{95\%}$(25 Hz) for fixed values of $\beta/H_{\rm pt}$ and $T_{\rm pt}$,  
                        and $v_{\rm w}= \kappa_{\phi} = 1 $. The dashed lines denote no sensitivity for exclusion. 
                        }
                \end{footnotesize}
                \label{tab:limits}
\end{center}
\end{table}
}

{
\begin{figure}[htb]
\begin{center}
\includegraphics[width=0.49\textwidth]{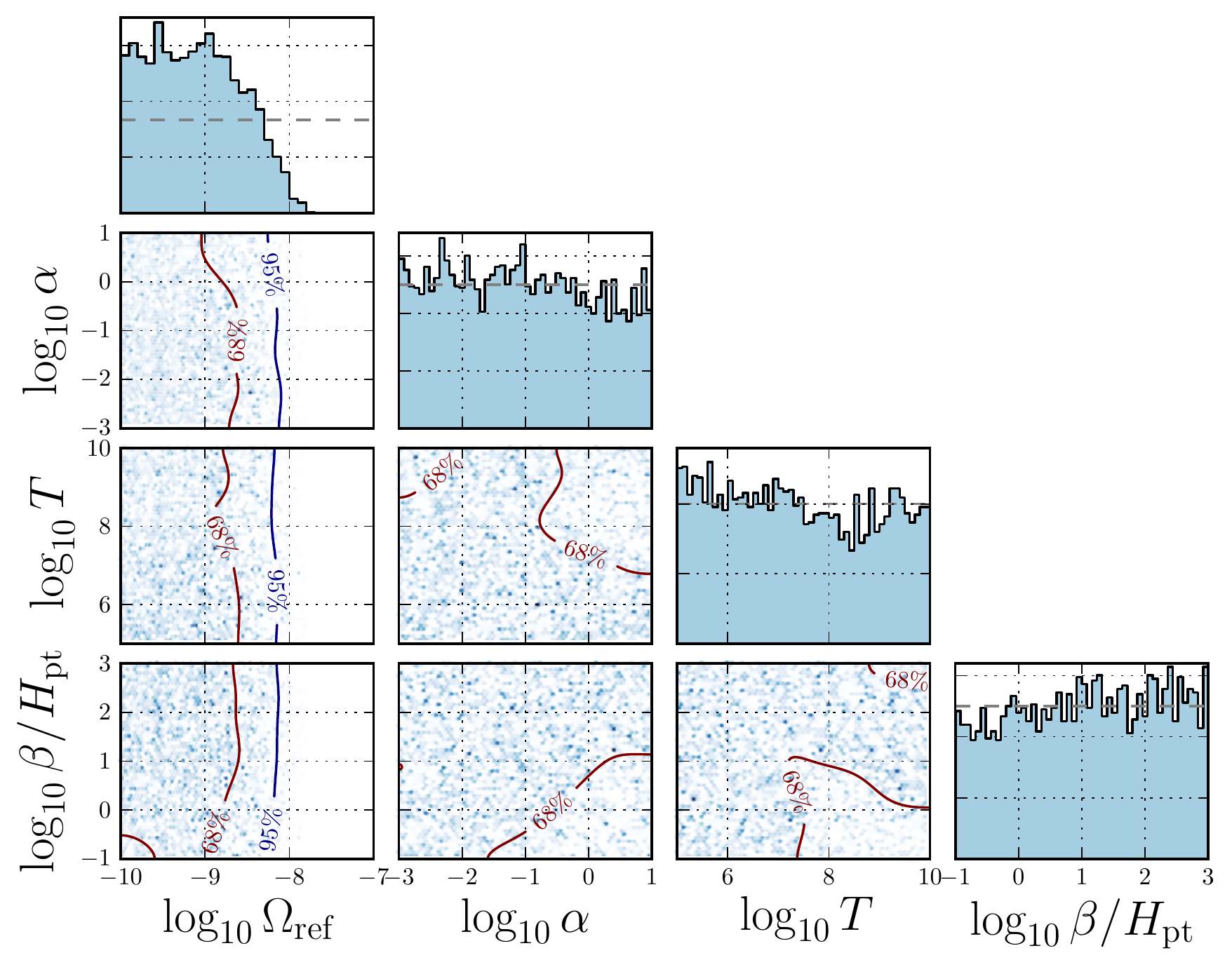}
\end{center}
\caption{\small
Posterior distributions for the CBC+FOPT search in the case of a phenomenological model with dominant sound wave  contributions ($v_{\rm w}= 1$),  as a function of 
$\log \Omega_{\rm ref}$ and the different parameters of the model.  The 68$\%$ and $95\%$ CL exclusion contours are shown. The horizontal dashed line in the 
posteriors indicate the flat priors used in the analysis.
}
\label{fig:SW}
\end{figure}
}

Similarly, in Fig.~\ref{fig:SW} we present the results for the CBC+FOPT hypothesis 
in which the sound waves dominate with $v_{\rm w}= 1$ and $\kappa_{\rm sw}$ a function of  $v_{\rm w}$ and $\alpha$.  The Bayesian 
analysis shows sensitivity at large values of $\alpha$  and $T_{\rm pt}$,  but  does not exclude  
regions in the parameter space at 95$\%$ CL.  
The analysis is then performed for given values of
$\beta/H_{\rm pt}$ and $T_{\rm pt}$ leaving $\alpha$ as a free parameter. As a result,  a 95$\%$ CL upper limit on 
$\Omega_{\rm sw}$(25 Hz)
of $5.9 \times 10^{-9}$ is obtained for $\beta/H_{\rm pt} < 1$ and $T_{\rm pt} > 10^8$ GeV. 
The analysis is repeated for models with reduced velocities of  
$v_{\rm w}= 0.9$,  $v_{\rm w}= 0.8$, and $v_{\rm w}= 0.7$, with Bayes factor $\log{\cal{B}}^{\rm CBC+sw}_{\rm noise} = -0.60$ and 
upper limit $\Omega_{\rm ref} \approx 5.9 \times 10^{-9}$, with no significant $v_w$ dependence.
In all studied cases, the models with reduced $v_w$ lead to significantly lower sound waves predicted energy densities, and  
with no 95$\%$ CL exclusions.


\sectionname{Conclusions}We have searched for signals from FOPTs in the early universe, potentially leading to a SGWB in the Advanced LIGO/Advanced Virgo frequency band. The analysis is based on the data from the three observation periods,  for which 
no generic stochastic signals above the detector noise has been observed.  

We use the results to deduce implications for models describing SGWB. We first consider a generic broken power law spectrum, describing its main features 
in terms of the shape and the peak amplitude. We place $95 \%$ CL upper limits simultaneously  
on the normalized energy density contribution from unresolved CBCs and a FOPT, $\Omega_{\rm CBC}(25\, \rm{Hz})=6.1 \times 10^{-9}$ and $\Omega_{\rm BPL}(25\,\rm Hz) = 4.4 \times 10^{-9}$, respectively.

The results are then interpreted in terms of  a phenomenological model 
describing contributions from  bubble collisions or sound waves, showing that  
the data can exclude a part of the parameter space at large temperatures. 
%
In a scenario in which bubble collision contributions dominate, with  $v_{\rm w}=1$ and $\kappa_\phi=1$, part of  the phase 
space with $T_{\rm pt} > 10^8$ GeV, $\alpha > 1$, and $\beta/H_{\rm pt} < 1$ is excluded at $95\%$ CL. 
For fixed values of $\beta/H_{\rm pt} = 0.1$, $1$ or $10$ and $T_{\rm pt} = 10^7, 10^8, 10^9$ or $10^{10} {\rm \ GeV}$,   the 95$\%$ CL upper limits on  $\Omega_{\rm coll}(25 \rm{Hz})$ vary in the range between
$ 4.0\times 10^{-9}$ and 
$ 1.0\times 10^{-8}$
which depends on the $\beta/H_{\rm pt}$ and $T_{\rm pt}$ values considered. 
In the case where sound waves dominate, several 
scenarios are explored considering different $v_w$. 
The data only shows a limited sensitivity,  and a  95$\%$ CL upper limit on  $\Omega_{\rm sw}(25 {\rm Hz})$
of $5.9  \times 10^{-9}$ is placed in the case of $v_{\rm w}=1$,  for  $\beta/H_{\rm pt} < 0.1$ and   $T_{\rm pt} > 10^8$ GeV.
%
Altogether, the results indicate the importance of using LIGO-Virgo GW data to place 
constraints on new phenomena related to strong FOPTs in the early universe~\cite{packagesnote}.


{
The authors would like to thank the LIGO-Virgo stochastic background group for helpful comments and discussions. In particular, the authors thank Patrick M. Meyers on his contributions to the parameter estimation analysis code. We thank Alberto Mariotti on his useful feedback on the draft. The authors are grateful for computational resources provided by the LIGO Laboratory and supported by National Science Foundation Grants PHY-0757058 and PHY-0823459. This paper has been given LIGO DCC number LIGO-P2000518.

A.R and M.M would like to thank O. Pujolàs for the motivation  and the fruitful discussions. 
This work was  partially  supported   by  the Spanish MINECO   under   the grants
SEV-2016-0588 and PGC2018-101858-B-I00,  some of which include ERDF  funds  from  the  European  Union.  
IFAE  is  partially funded by the CERCA program of the Generalitat de Catalunya. 
K.M. is supported by King's College London through a Postgraduate International Scholarship. M.S. is supported in part by the Science and Technology Facility Council (STFC), United Kingdom, under the research grant ST/P000258/1.
H.G. is supported by the U.S. Department of Energy grant No. DE-SC0009956.
F.W.Y. and Y.Z. are supported by the U.S. Department of Energy under Award No. DE-SC0009959.
}



\end{document}